# Structure and Substructure of Galactic Spheroids
**Astro2010 Science White Paper**

Aaron J. Romanowsky
UCO/Lick Observatory
(831) 459-2526
romanow@ucolick.org

Jean P. Brodie, UCO/Lick

James S. Bullock, UC Irvine

Robin Ciardullo, Penn State

Puragra Guhathakurta, UCO/Lick

Loren Hoffman, Northwestern

Knut A. G. Olsen, NOAO

Joel R. Primack, UC Santa Cruz

Glenn van de Ven, IAS

## Summary

The full spatio-chemo-dynamical structure of galaxies of all types and environments at low redshift provides a critical accompaniment to observations of galaxy formation at high redshift. The next decade brings the observational opportunity to strongly constrain nearby galaxies' histories of star formation and assembly, especially in the spheroids that comprise the large majority of the stellar mass in the Universe but have until now been difficult to study. In order to constrain the pathways to building up the spheroidal "red-sequence", various standard techniques in photometry and spectroscopy, particularly with resolved tracer populations like globular clusters and planetary nebulae, can be scaled up to comprehensive surveys as improved wide-field instrumentation is increasingly available. At the same time, progress in adaptive optics on giant telescopes could for the first time permit deep, resolved photometric and spectroscopic analysis of large samples of individual stars in these systems, thereby revolutionizing galaxy studies. Strong theoretical support is needed in order to understand the new observational constraints via detailed modeling and self-consistent simulations of star and galaxy formation throughout cosmic time.

## 1. Introduction

Nearby galaxies, as detailed end-products of the variety of observed high-redshift evolutionary processes, are a benchmark for theories of structure formation. A treasure trove of information about the evolutionary history of galaxies is encoded in the full phase space distribution of their observable constituents, especially the positions, velocities, ages, and metallicities of their stars. Neither observations nor simulations have been able to map out these basic properties in any reasonable detail, but the next decade heralds the opportunity to do so, providing a profound breakthrough for understanding galaxies.

The goal here is **holistic portraits of galaxies of all types and environments, from central black hole to outer dark halo**. This work is being realized in our own Milky Way (via SDSS, SEGUE, RAVE, GAIA, etc.) and in the dwarf and late-type galaxies of the Local Group and just beyond. However, the **massive spheroids**—dynamically hot stellar bulges and halos generally found in galaxies more luminous than $\sim 10^9$ $L_\odot$—have until now been largely out of reach. These spheroids are critical to understanding structure formation since they harbor the large majority of stellar mass in the present-day Universe [1], but unfortunately the nearest unobscured examples are at ~10 Mpc distances.

Attention in the last decade has increasingly focused on the "red-sequence" of massive spheroidal galaxies (ellipticals and lenticulars), which as hot systems of gas and old stars may represent a very different mode or phase of galaxy evolution from the star-forming "blue cloud" of late-types, and from the dwarf spheroidals. There have long been straightforward postulates for forming these galaxies, such as in mergers or by quenching of star formation, but the accumulating statistical data on the broad properties of galaxies over a range of redshifts have yielded puzzles about their sizes, masses, etc. [2,3]. If the overarching hierarchical clustering paradigm is correct, additional transforming mechanisms must be at work. Deconstructing these systems at low-redshift allows us to sift for clues about their formational histories, particularly in pursuit of mechanisms such as minor mergers that are very difficult to see in action at high redshift.

**Reconstructing the build-up of the red sequence** is a grand theme incorporating a wide variety of subplots, while requiring an assortment of both technical breakthroughs and incremental advances. Below we sketch out three key science questions for the next decade, along with the essential contributions that could be made at low redshift, and finally the potential of specific programs to provide the necessary progress. The three themes unite in the fundamental quest for the **joint spatio-chemo-dynamical structure of galaxies**.

## 2. Star formation histories

The present-day populations of stars in galaxies comprise a fossil record of their star formation (SF) histories, such that the timing and duration of major SF episodes may be recovered from the distributions of stellar ages and chemical abundances. The physics of star formation across cosmic time, including positive and negative feedback effects and the dependencies on galaxy mass and environment, is a major area of theoretical uncertainty in need of stronger constraints. The most massive spheroidals are thought to contain some of the oldest populations of stars in the Universe, formed early on in starbursts that may now be glimpsed with high-redshift observations [6,7]. If this scenario is true for the bulk of the stars in these galaxies, there still remain many uncertainties about SF epochs and timescales, chemical enrichment, and later stellar acquisitions via galaxy mergers and minor episodes of internal SF—questions that could in principle be answered by dissection of the complete stellar populations records in nearby galaxies.

The traditional approach to stellar populations in massive spheroidals is integrated light photometry and spectroscopy, in the latter case often confined to the very central galactic regions which may not be at all representative of the global properties, and in both cases subject to degeneracies in inferring ages and metallicities separately, and disentangling multiple populations. There have been gradual improvements in the quality and spatial extent of the observations as well as in the modeling [8,9,10], with results on age and metallicity gradients with radius that may be readily compared with detailed simulations of galaxy formation [11]. Steady observational and theoretical progress in this area should continue in the next decade.

Integrated light spectroscopy does not extend much beyond an effective radius ($R_{\text{eff}}$), into the outer galactic regions that are interesting for probing populations from accreted satellites and for testing theoretical predictions for radial gradients. Instead, globular clusters (GCs) and planetary nebulae (PNe) can be used as bright tracer objects, providing various age and abundance markers. As discrete objects representing single populations, they can also be invaluable for tracing secondary subpopulations [12,13]. For example, the metal-poor GCs found in all large galaxies probe the faint underlying metal-poor halo stars (and an important early stage of star formation), which even in nearby M31 were not discovered directly until very recently [14,15]. Increasing observational power in the next decade should enable the extensive photometric and spectroscopic data sets needed to make GCs and PNe practical tools for tracing star formation in many galaxies.

The most accurate way to study stellar populations is by resolving their color-magnitude diagrams (CMDs) as deeply as possible, yielding basic age and metallicity information. Because of distance and crowding effects, this technique has so far been mainly applied to very nearby dwarf galaxies and late-type galaxy disks and halos [14,15,16]. With the dramatic observational advances that may be feasible in the next decade, such **resolved studies could be extended to the bulges of galaxies as far as the Virgo cluster** (15 Mpc; Fig. 1), which contains a rich population of massive early-types. Spectroscopy of resolved stars would provide different populations indicators that could combine with CMD information to strengthen the overall constraints. The resolved studies in these galaxies would serve as **calibrations for studies of much more distant galaxies using integrated stellar light, GCs, and PNe**.

The myriad forthcoming constraints on stellar populations in galaxies of all types must go hand-in-hand with progress on the theoretical front to incorporate chemical processes in models of galaxy formation in a cosmological context. Such **chemodynamical simulations** would include realistic physical recipes for star and cluster formation, feedback, chemical enrichment, and mass assembly, with resulting star and galaxy properties that may be compared in detail to the observations at both high and low redshift.

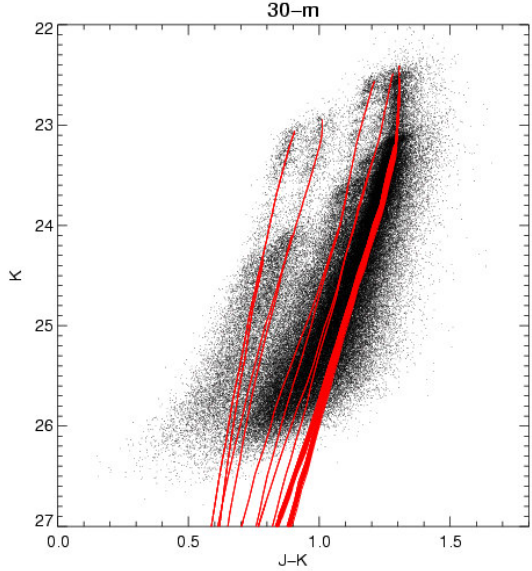

**Fig. 1.** Simulated color-magnitude diagram of ~$10^5$ resolved red giant branch (RGB) stars in NGC 3379 ($L^*$ early-type galaxy at 10 Mpc), observed in a 5 arcsec field at 1 $R_{eff}$ with a diffraction-limited 30-m telescope. The red lines show the input stellar populations models (constant metallicity tracks with [M/H] ~ −1.5, −1.0, −0.5, −0.2, +0.1, and ages ~9-11 Gyr) while the black dots show the recovered data. The magnitude limit (10% photometric accuracy) of $K$ ~ 25.7 is set by crowding, and improves to $K$ ~ 28.5 at 3 $R_{eff}$, outside of which it is limited to the same magnitude by sensitivity. In systems out to ~15 Mpc, the distribution of metallicities throughout most of the galaxy can be recovered, along with some age information [17]. In the galaxy halos, horizontal branch stars with young to intermediate ages would be detectable. For resolved spectroscopy, the crowding limits are the same, with sensitivity and instrumental considerations becoming dominant. Obtaining the brightest RGBs out to ~15 Mpc could be feasible with bluer AO functionality (*I*-band), which would also enhance CMD studies. The spectra of many stars from a CMD sub-region could be stacked for abundance measurements (e.g, Ca) at higher $S/N$.

## 3. Mass inventories

Another cardinal question about a galaxy is its distribution of mass: how much and where are its stars, gas, black holes, and **dark matter** (DM)? We have a good idea of how most of the *stellar luminosity* is distributed in galaxies, but there remain notably large uncertainties about the DM on galactic scales. One problem is in the central regions, which are important for probing DM properties and the interplay between baryons and DM during galaxy assembly: cold dark matter (CDM) theory predicts central DM "cusps" that could be further altered by the back-reaction of baryon collapse and outflows. The dynamical mass is relatively easy to determine in these regions (see below), but the stellar mass-to-light ratios ($M/L$) are not known with enough precision to decompose the total mass into the stellar and DM contributions—particularly in the case of high surface brightness early-type galaxies where the stars dominate the central mass. This problem may be solved before long in Local Group dwarf galaxies [18], but it is important to determine the distribution of DM for the entire range of galaxies and environments.

Determining the **stellar $M/L$ including its spatial variations in galaxies** will be helped by ongoing improvements in stellar evolutionary models which refine the accuracy of the mass corresponding to the directly observable stellar populations, along with observational advances as discussed in the previous section. However, a major residual source of uncertainty is the amount of "hidden" mass in faint and low-mass objects, represented by the initial mass function (IMF): although the IMF is known reasonably well in the Milky Way, it could be very different in other galaxies depending on the initial conditions. This question can be addressed by various programs tangentially related to early-type galaxies, such as modeling the formation and disruption of stars and star clusters in different environments, along with comparisons of field and cluster CMDs, and dynamical measurements of the masses of face-on galactic stellar disks.

In the outer regions of galaxies, DM is dominant and the stellar $M/L$ issues are less important, but finding and exploiting mass tracers becomes a challenge. Late-type galaxies contain gas disks that constrain the total mass, but these do not provide much leverage on the three-dimensional shape of the gravitational potential. Early-type galaxies lack even the gas disks. Gravitational lensing (weak and strong) does yield important mass constraints, particularly over a range of redshifts, but cannot deliver detailed mass distributions for arbitrarily selected individual galaxies. Mass inferences with thermal X-ray gas are limited to the very brightest ellipticals. As with stellar populations studies, the traditional approach is to use integrated-light spectroscopy, extracting detailed stellar kinematical information and then using high-powered dynamical models to estimate the mass [19].

Integrated light stellar kinematics is becoming efficient enough to push out to ~ 3 $R_{eff}$ [20,21], but to really tackle the DM problem requires constraints out to ~ 8 $R_{eff}$ (~ 30 kpc) which is the typical predicted DM halo scale radius for an $L^*$ galaxy in ΛCDM. This radial extent is not usually feasible even with gas-rich spirals, so the only remaining generic alternative is to obtain velocities of large numbers of resolved tracer objects. Traditional work has employed GCs and PNe as useful tracers in nearby early-type galaxies of ~ $L^*$ and brighter. Samples of ~$10^2$-$10^3$ velocities are currently feasible, and with future instrumentation could be increased to much larger and more distant samples of galaxies. However, the dynamical analyses still require some degree of simplifying assumptions about orbit types, mass profile, and/or geometry, along with auxiliary mass constraints; ~$10^4$ velocities are needed for the **fully general reconstruction of mass distributions**.

As introduced in the previous section, advances in adaptive optics (AO) on giant telescopes raise the possibility of studying the **kinematics of individual stars** for the first time outside the Local Group. Using RGB stars as tracers of old stellar populations, the disks, bulges, and halos of all galaxy types can be probed, with samples of perhaps even ~$10^6$ velocities per system. With a representative sample of galaxies studied in such detail, the systematic uncertainties in using GCs and PNe can be more fully understood, allowing these brighter tracers to be used with greater confidence in galaxies at much greater distances.

The dynamical models for extracting mass profiles must be suited to such a wealth of data, allowing for fully triaxial systems whose very outer parts may not be in equilibrium. The theoretical predictions must also be improved, as it is already apparent that DM-only simulations do not reproduce the variety of mass profiles found in the Universe [22], and that it is crucial to understand the interplay of baryons and DM [23].

Another major connected theme is the **supermassive black holes** (SMBHs) that are pervasively found in the centers of giant galactic spheroids and are deeply connected to the processes of galaxy formation. The next decade should see further progress in studying SMBHs using stellar dynamics, which couples to global studies of spheroids since knowing the mass distribution at large radius can be critical for accurate SMBH mass inferences [24], and the SMBHs can themselves significantly affect the stellar orbits (the theme discussed next).

## 4. Dynamical structure and substructure

The six-dimensional position-velocity phase space structure (or distribution function, DF) of a galaxy is a powerful record of its assembly history. Processes such as galaxy mergers, stream-fed gas accretion [25], and quiescent disk-mode SF should all leave distinct dynamical signatures on the stellar populations. For example, decades of detailed observations of early-type galaxies culminating in the SAURON/Atlas3D surveys have resolved these systems into two basic classes: the faint, disky "fast rotators", and the bright, boxy, "slow rotators" [26,27]. The fast rotators seem readily explained as forming in major mergers of gas-rich galaxies, while the intrinsic shapes and dynamics of the slow rotators are as yet difficult to reproduce theoretically but may involve rapid multiple mergers at high redshifts [28,29]. The connections between these early-type galaxies and the bulges and pseudo-bulges of late-type galaxies are also unclear [30].

Further information on the pathways to galaxy formation can be obtained by extending kinematics observations well outside of $R_{eff}$ into the realm of **galaxy halos**. However, even the most basic information about stellar orbits in halos is so far undetermined, with recent suggestions of radial anisotropy in fast rotators and tangential in slow rotators [22,31]; the halos should also be reservoirs of angular momentum— a fundamental parameter which again is almost totally uncharted in massive spheroids. Both observations and simulations also now suggest that the halo dynamical structure in early-type galaxies can be radically different from the central regions (Fig. 2; see also [33]). This core-halo "decoupling" could be used to infer the different baryonic physics and merger parameters that formed individual galaxies. Other contributing information could come from detailed study of unrelaxed outer-halo substructures such as shells, streams, and moving groups of stars, PNe, and GCs [34,35], transitioning at the outermost radii into the exchange of material between a galaxy and its environment, as part of the broader theme of **intracluster and intragroup light** (ICL)—which benefits from many of the same techniques used in studying galaxy halos.

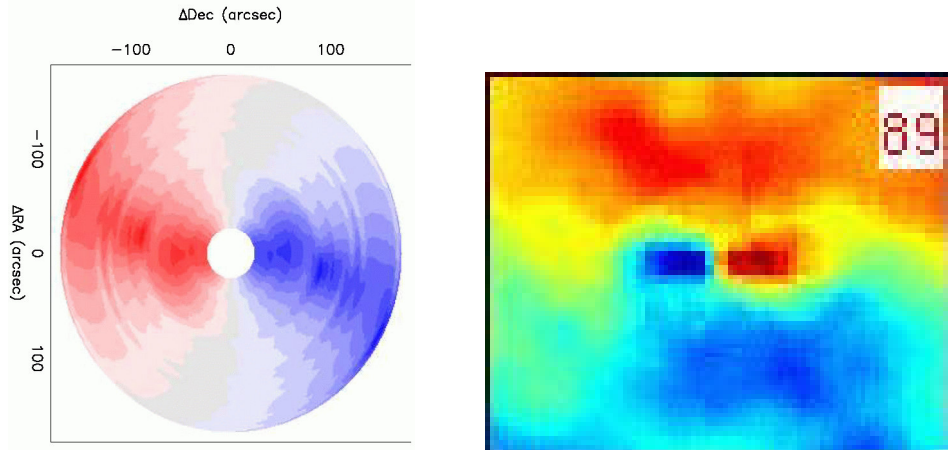

**Fig. 2.** Stellar rotation field of a fast-rotator early-type galaxy, out to ~3 effective radii ($R_{eff}$), where colors represent red- and blue-shifts. *Left*: Observations of NGC 4494 using Keck/DEIMOS in a novel mode of "stellar kinematics with multiple slits (SKiMS)" [21]. *Right:* Simulated remnant of a gas-rich equal-mass galaxy merger [32]. In both cases, the rotation axis twists strongly outside ~1 $R_{eff}$: a core-halo decoupling that is not readily apparent in the photometry. The central axisymmetric regions as typically probed by SAURON may originate in gas dissipational processes and the birth of new stars, with the outer regions corresponding to the triaxial remnant of old stars from the merged progenitor galaxies.

To pursue these issues requires the same extensive kinematical data sets and related modeling tools discussed in the context of mass inferences. An additional goal is to determine how the DF depends on age-metallicity subpopulations, providing the ultimate tool for extracting a galaxy's **combined history of assembly and star formation** (already an established enterprise in the Milky Way). Very little work has been done along these lines in early-type galaxies because of the challenges in modeling multiple populations in integrated stellar light [36]. Even with a single population, no more than a few higher-order velocity moments may be measured, limiting the details of the orbital structures that may be extracted. However, a tremendous window of opportunity is opened up by the use of discrete tracers, where it becomes fairly straightforward to separate subpopulations, and where sufficient numbers could resolve subtle dynamical substructures. The most potent application of this approach would be the use of resolved stars, whose photometry and spectroscopy out to the Virgo cluster could be feasible in the next decade, as previously discussed. Again, the resolved stellar analyses would also serve to calibrate the use of integrated light, GCs, and PNe as tracers in denser and more distant systems.

## 5. Programmatic opportunities

Many of the observational and theoretical tools required to advance our understanding of massive spheroids are well-established but can become much more effective with steadily increasing instrumentational and computational power in the next decade. On the other hand, a qualitative breakthrough in nearby galaxy studies could occur with the use of giant-telescope AO to obtain resolved stellar photometry and spectroscopy in systems far beyond the Local Group. Given the multiple variables involved, it is difficult to forecast the exact AO capabilities that could be developed, but simulations and extrapolations provide a reasonable guide.

The main unifying observational theme here is the efficient, extensive use and increased role of **resolved photometric and spectroscopic tracers** in all types of galaxies, including GCs, PNe, and giant stars. Most of the proposed major optical and infrared facilities for the next decade would be highly useful for this work, as long as **general-purpose capabilities and open access** are maintained. Especially noteworthy progress can be made via benchmark comprehensive surveys of nearby spheroids, given the allocation of resources to **integrated multi-wavelength surveys** to efficiently provide uniform data sets and analyses (cf. SAURON and ACSVCS [27,37]).

The initial key observational need is wide-field ground based photometry and multi-object spectroscopy (MOS) on 10m telescopes, in order to carry out the dynamics studies with integrated stellar light (via SKiMS), GCs, and PNe, and populations analyses using stellar light and GCs. There are effective facilities currently available (e.g. Subaru/Suprime-Cam, Keck/DEIMOS; [31]) but progress is slow due to the sheer number of spectra required with long integration times. This leads to problems of incompleteness and inhomogeneity in individual galaxies studied, and low numbers of galaxies surveyed. The situation could be dramatically improved with the next round of wider-field instruments (e.g., Subaru/Hyper-Suprime-Cam, Subaru/WFMOS). In this case, the MOS $S/N$ would still limit the observations to $\sim 10^2$-$10^3$ spectra per galaxy, and galaxy distances to ~25 Mpc. The advent of a GSMT with a wide-field seeing-limited optical spectrograph (e.g., GMT/GMACS, TMT/WFOS) would be a great benefit, increasing the range to ~50 Mpc (an eightfold increase in the survey volume) and the number of tracer objects in the nearer galaxies by an order of magnitude [38]. This facility would also enable decisive PN-based populations work to finally be practical at Virgo distances (using the faint [O III] line at 4363Å to probe progenitor age and metallicity). Finally, the LSST would be a peerless boon for studying the GC systems of galaxies across the entire sky, with co-added *griz* imaging extending the 50% completeness range to ~50 Mpc. It could also conceivably be used to trace resolved RGB stars in the extreme outer regions of galaxies out to ~10 Mpc (e.g., at *V*-band surface brightness levels fainter than ~29 mag/arcsec$^2$). In parallel to the ground-based advances, space-based (ACS, WFC3, JWST) work could progress with traditional deep populations studies of resolved stars and GCs. **General observer access to JDEM beyond the primary mission** would permit spectacular improvements in this area, with $\sim 10^3$ times greater areal coverage per galaxy.

In the era of grand achievements with costly projects and surveys, a niche should be preserved for **modest-priced, specialized instruments and novel techniques** which can still have a major impact on focused science areas. Current examples include SAURON, the PN.S, the Burrell Schmidt, and BlackBird Remote Observatory, all on 4-meter and smaller telescopes [27,33,35,39]. Future possibilities include the use of tunable narrow-band filters that could obtain spectrophotometry of integrated light, GCs, and PNe in single observing runs of galaxies [40]. Previous standard techniques could be enhanced into large efficient surveys, e.g. the sample of well-determined photometric characteristics of nearby galaxies (isophote shapes, color gradients, etc.) is surprisingly small but would not require a large telescope to remedy. Also the SAURON/Atlas3D surveys of two-dimensional kinematics and abundances could be extended to larger radii and multiple population tracers with new **wide-field IFUs** (cf. HET/VIRUS, Keck/CWI).

The centerpiece of potential future progress on nearby spheroids is AO on GSMT. With facilities such as TMT/IRIS or GMT/NIRAOI, much deeper CMDs than possible with HST or JWST could be constructed (see Fig. 1). **Extending AO capabilities into the optical** as far as possible would be valuable for stellar populations constraints, and critical for enabling extensive **metallicity and kinematics studies of individual stars** using the *I*-band calcium triplet. Note that the large-aperture telescope is needed for sensitivity but high-Strehl diffraction-limited AO is not necessary; effective PSFs of ~30 mas would be adequate for coping with crowding limitations, and one could consider relatively wide-field MOS with "enhanced-seeing" (cf. TMT/IRMS). Such studies in galaxies out to Virgo (15 Mpc) distances would act as detailed references for studies of GCs, PNe, and integrated light which then could stretch out to Coma cluster (100 Mpc) distances, encompassing the full range of galaxy environments including voids, active mergers, compact and fossil groups, and massive clusters. Note that spheroid science need not await the full GSMT capabilities, as 10-m AO and first-light GSMT instruments could probe GCs in crowded regions, and RGBs in nearby spiral bulges and NGC 5128, before advancing to 15 Mpc.

**A new generation of dynamical models** will be required to extract mass distributions and orbital properties from the next decade's anticipated data bonanza. The models can rely on well-established techniques such as Schwarzschild orbit modeling, DF basis functions, or particle-based methods [41,42], modified and scaled up accordingly. With respect to modeling discrete velocity data, $\sim 10^3$ tracers can constrain the mass of an axisymmetric system to $\sim 10\%$ precision, with the internal structure somewhat less well determined [43]. Allowing for more generalized triaxial structure introduces more free parameters but also provides more signatures in the data such as kinematic twisting which help determine the geometry [44]; $\sim 10^4$ tracers are probably enough, while pinning down the orbital details of multiple subpopulations might require $\sim 10^5$. An additional novel approach might be developed as a hybrid between traditional dynamical modeling and galaxy formation simulations, whereby the simulations might be evolved to a snapshot that matches the data, perhaps using a forcing function or else via iterative genetic techniques [45].

The third leg of the spheroids study stool is galaxy formation theory, which is a large enterprise with many faces. The ultimate goal of simulating the universe over all relevant scales (stars to CMB) is a long way off, but gradually approached via increasing computational power and improved understanding of various physical processes. Some basic complementary approaches include numerical simulations of galaxy formation at different levels of detail, semi-analytic models which use simplified prescriptions for unsimulated processes, and analytic theory to help interpret and refresh the other approaches. The details of DM assembly is virtually a "solved" problem, but the baryonic investigations are only beginning, requiring **high-resolution hydrodynamic simulations**, including cases in a full cosmological context as well as toy models and resimulations of smaller regions for finer resolution on galactic scales [28,29]. Some current areas of interest are how feedback occurs and star formation is regulated in the galactic environment. In this context, theoretical understanding of massive spheroid evolution will progress by simulations and analyses that are tailored to the relevant observations (e.g., studying the dynamical properties of simulated galaxy mergers). One action that would facilitate progress would be the establishment of a **national theory center in galaxy formation and/or computational astrophysics**.